\documentclass[twocolumn,amssymb, nobibnotes, aps, prd,amsmath,mathtools, nofootinbib,pgfmath,pgffor,superscriptaddress,floatfix]{revtex4-1}
\usepackage{graphicx}
\usepackage{ulem,amsthm}
\usepackage{enumitem}

\makeatletter
\def\qrr@split@result#1 #2\@qrr@split@result{\edef\erfInput{#1}\edef\erfResult{#2}}
\newcommand*{\gnuplotErf}[2][\jobname.eval]{%
    \immediate\write18{gnuplot -e "set print '#1'; print #2, erf(#2);"}%
    \everyeof{\noexpand}
    \edef\qrr@temp{\input{#1} }%
    \expandafter\qrr@split@result\qrr@temp\@qrr@split@result
}
\makeatother

\usepackage{caption,color}
\usepackage[caption=false]{subfig}
\usepackage{multirow}

\usepackage{hyperref}
\hypersetup{colorlinks,citecolor=blue}

\usepackage[dvipsnames]{xcolor}

\begin{document}

\title[Firewall black holes and echoes from an action principle]{Firewall black holes and echoes from an action principle}

\author{Jahed Abedi}
\email{jahed.abedi@uis.no}
\affiliation{Department of Mathematics and Physics, University of Stavanger, NO-4036 Stavanger, Norway}

\begin{abstract}
It is often said that there is no gravity theory based on local action principles giving rise to firewall black hole solutions. Additionally, Guo and Mathur 2022 \cite{Guo:2022umn} have cast doubt on the observability of firewall echoes due to closed trapped surface produced by backreaction of macroscopic in-falling wave packets. In this paper, we bring Einstein-Maxwell-Dilaton action as a toy model that serves as counterexample to these assertions. Actions with Maxwell and dilaton fields emerge from several fundamental theories, such as the low energy limit of (super) string theory or Kaluza-Klein compactifications. In these systems, the black hole solution has two curvature singularities. We will show that the outer singularity inside the event horizon can cause significant change to the outside, close to the extremal limit, making a macroscopic reflective barrier near the event horizon that would lead to ``observable'' gravitational wave echoes in this toy model.
Additionally, we also call into question the argument by Guo {\it et al}. 2017 \cite{Guo:2017jmi} claiming that a very small fraction of the backscattered photons will be able to escape back to infinity from the firewall using these black holes as counterexample.
\end{abstract}

\maketitle

\section{Introduction}
Black holes are potential gateways to groundbreaking discoveries. Black hole (BH) astrophysics has undergone an observational renaissance in the past six years. Notably, the observation of gravitational waves has provided an exciting new window to probe as close as possible to the event horizon of observed binary BH mergers \cite{Cardoso:2016rao}.
With these observations, the closer to the event horizon we probe, the place where we expect to see exciting and very nontrivial behavior of quantum gravity, the higher energy physics we achieve. As several approaches suggest evading the information paradox \cite{PhysRevD.14.2460} by replacing the region around the event horizon by a firewall or exotic compact object (ECO) makes it a potential target for discoveries on departures from general relativity. One intriguing question is how to keep both the equivalence principle and quantum mechanics and still find BHs from local action principles that carry a firewall and/or ECO. In this paper we bring a toy model to answer this question.
Another intriguing question is whether having a firewall covered by an event horizon can make observable echoes. Our answer to this question is affirmative. This paper also brings Einstein-Maxwell-Dilaton action as an example, to answer several essential questions which are arising in the context of observability of gravitation wave echoes, firewall, and ECO \cite{Guo:2017jmi,Guo:2022umn}.

The possibility of observing gravitational wave echoes has led to several observational searches \cite{Abedi:2021tti,LIGOScientific:2021sio,Westerweck:2021nue,LIGOScientific:2020tif,Abedi:2020sgg,Abedi:2020ujo,Wang:2020ayy,Tsang:2019zra,Holdom:2019bdv,Salemi:2019uea,Uchikata:2019frs,Abedi:2018npz,Abedi:2018pst,Nielsen:2018lkf,Lo:2018sep,Conklin:2017lwb,Westerweck:2017hus,Abedi:2017isz,Ashton:2016xff,Abedi:2016hgu} with positive \cite{Abedi:2021tti,Holdom:2019bdv,Uchikata:2019frs,Abedi:2018npz,Conklin:2017lwb,Abedi:2016hgu}, mixed \cite{Salemi:2019uea,Nielsen:2018lkf,Westerweck:2017hus}, and negative \cite{LIGOScientific:2021sio,Westerweck:2021nue,LIGOScientific:2020tif,Wang:2020ayy,Uchikata:2019frs,Tsang:2019zra,Lo:2018sep} results.

This system of Maxwell and dilaton fields coupled to gravity  emerges from a number of more fundamental theories. Notably the (super) string theory at low energy and Kaluza-Klein compactifications leads to such actions, which  have been studied for a long time \cite{ Gibbons:1987ps, PhysRevD.43.3140, Holzhey:1991bx, Koga:1995bs,Abedi:2013xua}.
Corresponding BHs and their evaporation have also been studied previously \cite{Abedi:2013xua,Holzhey:1991bx}. The behavior of the theory in the range of $\alpha\geq1$ is significantly different and shows unexpected features \cite{Abedi:2013xua,Holzhey:1991bx}.
Since this BH poses an electric charge, it makes more sense to study the charged particles scattering in this background. Here we chose fermions and scalars.
The greybody factors, evaporation and evolution of dilaton BHs for fermions studied by Abedi {\it et al}. \cite{Abedi:2013xua}. Consideration of the backreaction draws a highly nontrivial picture revealing a new phenomenon of evaporating to an extremal limit as the fate of certain dilaton BHs \cite{Abedi:2013xua}. Holzhey {\it et al}. \cite{Holzhey:1991bx} derived the potential barrier for scalars which for $\alpha>1$ strongly impedes the particle radiation to the extent that it may stop it. In contrast, Koga {\it et al}. \cite{Koga:1995bs} showed by numerical computation that Hawking radiation wins over the barrier and the dilaton BH does not stop radiating, despite the fact that the potential barrier becomes infinitely high. Finally, considering the backreaction and adiabatic approximations, Abedi {\it et al}. \cite{Abedi:2013xua} have shown that for fermions the potential barrier stops the BH from radiation at the extremal limit although it has a divergent Hawking temperature. Considering the next order of dynamical effect as backreaction changes the fate of the BH in which it becomes the key factor when the BH evolves toward the extremal limit. In other words, the nontrivial spacetime around the event horizon of this BH extinguishes the Hawking radiation.
Certain results on the scattering parameters of the Dirac field such as quasinormal frequencies or decay rates in the background of dilaton BH \cite{Gibbons:2008rs,Nakonieczny:2011bs,Ong:2019glf} are also presented.


\section{Dilaton black holes}
Einstein-Maxwell-Dilaton gravity with the dimensionless dilaton coupling constant $\alpha$ is given as follows:
\begin{equation}
S=\int d^{4} x \sqrt{-g}[R-2(\nabla\phi)^{2}+e^{-2\alpha\phi}F^{2}] \label{eq2.1}.
\end{equation}
Here $\phi$ is dilaton the field, and $F^{2}=F_{\mu\nu}F^{\mu\nu}$. The Einstein-Maxwell-Dilaton equations of motion of this gravity are,
\begin{equation}
\nabla_{\mu}(e^{-2\alpha\phi}F^{\mu\nu})=0,
\end{equation}
\begin{equation}
\partial_{[\rho}F_{\mu\nu]}=0,
\end{equation}
\begin{equation}
R_{\mu\nu}=e^{-2\alpha\phi}(-2F_{\mu\rho}F^{\rho}_{\nu}+\frac{1}{2}F^{2}g_{\mu\nu})+2\partial_{\mu}\phi\partial_{\nu}\phi,
\end{equation}
\begin{equation}
g^{\mu\nu}\nabla_{\mu}\nabla_{\nu}\phi=\frac{1}{2}\alpha e^{-2\alpha\phi}F^{2} \label{eq2.5}.
\end{equation}

Static spherically symmetric BH solutions of this metric are given by\cite{Gibbons:1987ps,PhysRevD.43.3140},

\begin{multline}
ds^{2}=(1-\frac{r_{+}}{r})(1-\frac{r_{-}}{r})^{\frac{1-\alpha^{2}}{1+\alpha^{2}}}dt^{2} \\ -\frac{dr^{2}}{(1-\frac{r_{+}}{r})(1-\frac{r_{-}}{r})^{\frac{1-\alpha^{2}}{1+\alpha^{2}}}}-r^{2}(1-\frac{r_{-}}{r})^{\frac{2\alpha^{2}}{1+\alpha^{2}}}d\Omega^{2} \label{eq6},
\end{multline}
with the Maxwell $A_{t}=-\frac{Q}{r}$, and dilaton fields $e^{2\alpha\phi}=(1-\frac{r_{-}}{r})^{\frac{2\alpha^{2}}{1+\alpha^{2}}}$.
The solution contains outer and inner horizons located at
\begin{equation}
r_{+}=M+\sqrt{M^{2}-(1-\alpha^{2})Q^{2}},
\end{equation}
\begin{equation}
r_{-}=\frac{1+\alpha^{2}}{1-\alpha^{2}}(M-\sqrt{M^{2}-(1-\alpha^{2})Q^{2}}),
\end{equation}
where M and Q are ADM mass and charge of the BH respectively.

The Hawking temperature of the dilaton BH is given by
\begin{equation}
T_{H}=\frac{1}{4 r_{+}}(1-\frac{r_{-}}{r_{+}})^{\frac{1-\alpha^{2}}{1+\alpha^{2}}}.
\end{equation}
This BH exhibits some interesting and unique thermodynamical properties \cite{Holzhey:1991bx,Koga:1995bs,Preskill:1991tb,Koga:1994np,Abedi:2013xua}. For $\alpha<1$, it is similar to the Reissner-Nordstr\"{o}m (RN) BH where the temperature approaches zero when it gets closer to the extremal limit. Nontrivial behavior occurs for $\alpha>1$ and $\alpha=1$. For $\alpha>1$, at the extremal limit the temperature diverges, while for $\alpha=1$ it converges to a finite  nonzero value $T_{H}=1/4\pi r_{+}$.

For nonzero $\alpha$ and for extremal BHs the angular factor in the metric (\ref{eq6}) and correspondingly its area vanish at the event horizon, and the geometry becomes singular, while no such singularity exists for RN BH ($\alpha=0$).

The surface area A of the BH and its entropy is given by the Bekenstein-Hawking formula,
\begin{equation}
S_{BH}=\frac{1}{4}A=\pi r_{+}^{2} \left( 1-\frac{r_{-}}{r_{+}} \right)^{\frac{2\alpha^{2}}{1+\alpha^{2}}}.\label{eq2.14}
\end{equation}

As obtained in \cite{Abedi:2013xua}, the peak of the effective potential for both scalars and fermions is approximately at
\begin{multline}
\frac{r_{\rm{max}}}{r_{+}}=1-\frac{1}{4} \left( \frac{3-\alpha^{2}}{1+\alpha^{2}} \epsilon - 2 \frac{1-\alpha^{2}}{1+\alpha^{2}} \right) \\
+ \frac{1}{4}\left( \left(\frac{3-\alpha^{2}}{1+\alpha^{2}} \epsilon - 2 \frac{1-\alpha^{2}}{1+\alpha^{2}} \right)^2 + 8\epsilon \right)^{\frac{1}{2}} \label{eq2.33},
\end{multline}
where we have $\epsilon=1-\frac{r_{-}}{r_{+}}$. 

For BHs with no charge ($\epsilon=1$) the location of maximum is at $r_{\rm{max}}=\frac{3}{2}r_{+}$ which is what we anticipated from the Schwarzschild BH. Increasing the charge changes the position of the peak, where its moving direction depends on the value of the coupling constant $\alpha$. For the BHs in the range $0\leq\alpha<1/\sqrt{3}$, with an increasing charge, the position of the peak moves away from the horizon approaching $r_{\rm{max}}\rightarrow\frac{2r_{+}}{1+\alpha^{2}}$ at the extremal limit when $\epsilon\rightarrow0$. For the particular value of $\alpha=1/\sqrt{3}$, the location of the peak remains fixed at $r_{\rm{max}}=\frac{3}{2}r_{+}$.
For the range $1/\sqrt{3}<\alpha<1$, by addition  of the charge, the location of peak retreats toward the event horizon and tends to $r_{\rm{max}} \rightarrow \frac{2 r_{+}}{1+\alpha^{2}}$ at the extremal limit. In the case of $\alpha\geq1$ which is the most interesting range where the location of the peak always stays in the range $r_{+}\leqslant r_{\rm{max}}\leqslant\frac{3}{2} r_{+}$, the near extremal BH acts like a firewall. When it approaches the extremal limit, the peak moves toward and finally touches the event horizon. At the limit $\alpha\rightarrow\infty$, the peak approaches $r_{\rm{max}}=\left(1+\frac{\epsilon}{2}\right)r_{+}$ and for $\alpha=1$, $r_{\rm{max}}\rightarrow(1+\sqrt{\frac{\epsilon}{2}})r_{+}$ 
 and the case of $\alpha>1$ has $r_{\rm{max}}\rightarrow\left(1+\frac{1}{2}\frac{\alpha^{2}+1}{\alpha^{2}-1}\epsilon\right)r_{+}$.

For fermions with charge $q$ and energy $\omega$ the approximate function of the peak in our region of interest $\alpha\geq1$ is as follows \cite{Abedi:2013xua}:
\begin{equation}
\left(V_{1,2}\right)_{\rm{max}} \simeq \frac{ \kappa^{2} }{ r_{+}^{2} \left(1-\frac{qQ}{\omega r_{+}}\right)^{2}  }, \ \ \ \ \ \ \ \ \ \ \ \ \ \ \ \ \ \ \ \ \ \ \alpha=1 \label{eq3.20},
\end{equation}
and
\begin{multline}
\left(V_{1,2}\right)_{\rm{max}} \simeq \frac{ \kappa \left(\kappa + \frac{1-\alpha^{2}}{1+\alpha^{2}}\right)}{ r_{+}^{2} \left(1-\frac{qQ}{\omega r_{+}}\right)^{2} \left[\frac{1}{2}\frac{\alpha^{2}-1}{\alpha^{2}+1} \left(1-\frac{r_{-}}{r_{+}}\right)\right]^{\frac{2\alpha^{2}-2}{\alpha^{2}+1}} },\\  \alpha>1\label{eq3.9},
\end{multline}
where ($\kappa=\pm1,\pm2,...$) are integer numbers.

For scalars the approximate function of the peak for $\alpha\geq1$ is given by \cite{Abedi:2013xua}
\begin{equation}
\left(V_{\eta}\right)_{\rm{max}} \simeq \frac{ (l+\frac{1}{2})^{2} }{ r_{+}^{2} }, \ \ \ \ \ \ \ \ \ \ \ \ \ \ \ \ \ \ \ \ \ \ \ \ \ \ \  \alpha=1, \label{eq3.24}
\end{equation}
and
\begin{equation}
\left(V_{\eta}\right)_{\rm{max}} \simeq \frac{ (l+\frac{1}{2})^{2} - \frac{1}{4}\left(\frac{1-\alpha^{2}}{1+\alpha^{2}}\right)^{2} }{ r_{+}^{2} \left[\frac{1}{2}\frac{\alpha^{2}-1}{\alpha^{2}+1} \left(1-\frac{r_{-}}{r_{+}}\right)\right]^{\frac{2\alpha^{2}-2}{\alpha^{2}+1}}}. \ \ \alpha>1 \label{eq3.17}.
\end{equation}

\begin{figure}[t]
\centering
\includegraphics[width=0.49\textwidth]{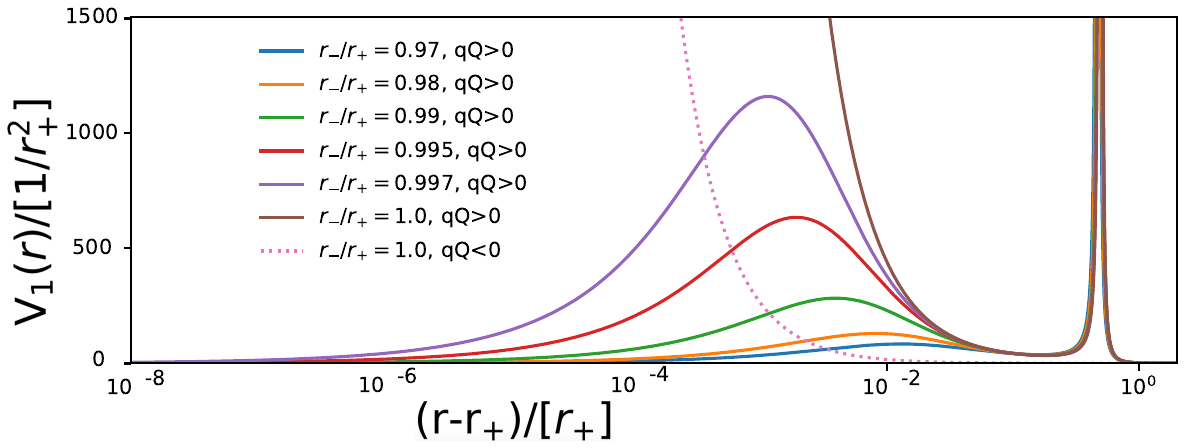}
\caption{\label{fig_1}Effective potentials (barrier on right) due to angular momentum (firewall) and electric potential (barrier on the left) with $\alpha=2$ and different values of charge (see the Appendix for details).  Here as we approach the extremal limit, the peak of the barrier grows and approaches the event horizon. At the extremal limit it touches the event horizon.}
\end{figure}

Figure \ref{fig_1} shows the plot of the potential barrier for fermions for the value of $\alpha=2$ ($\alpha>1$) in near extremal and extremal which are our regime of interest in this paper. For details of the behavior of these potentials we refer to \cite{Abedi:2013xua}.

Since the charge of the emitted particle appears as $\Phi_{H}=\frac{qQ}{r_{+}}$ in the denominators of the maximum of the potentials (\ref{eq3.20}) and (\ref{eq3.9}), the peak is higher for the case when the emitted charge and BH charge are the same. Surprisingly, particles with opposite charge leave the BH easily\footnote{Note that the total particle creation rate out of vacuum  dominates this effect \cite{Abedi:2013xua}, and the BH loses charge by time.}. As is shown in Fig. \ref{fig_1}, we also see that these potentials are divergent due to the same electrical potential term $\omega=q\Phi_{H}=\frac{qQ}{r_{+}}$. So these particles hardly escape the BH. We will get back to this point in subsequent discussions.

The effective potential is very unique for the dilaton BHs. Unlike the standard BHs and the case $\alpha<1$ (where the peak of the potential remains finite and tends to zero when we approach the horizon), for $\alpha=1$ the maximum (\ref{eq3.20}) does not vanish at the horizon limit, which replicates a soft firewall. For $\alpha>1$, (\ref{eq3.9}) becomes very large and divergent and blows up at the horizon, which replicates a hard firewall which is shown in Fig. \ref{fig_1}.

\begin{figure}[t]
\centering
\includegraphics[width=0.45\textwidth]{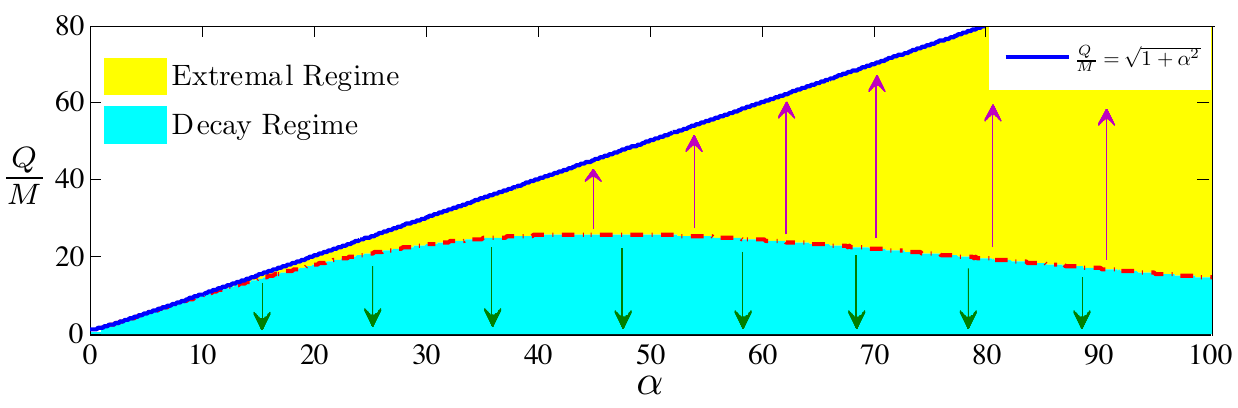}
\caption{\label{fig_5}Transition line, extremal line, extremal regime, decay regime, and direction of evolution and fate of dilaton BHs (see the Appendix for details).}
\end{figure}

These BHs evolve into two possible final states \cite{Abedi:2013xua}: spontaneously evaporating toward the extremal limit, or complete evaporation. The boundary (transition line) of the separation of these two conditions is specified in the $(Q/M,\alpha)$ plane (Fig. \ref{fig_5}). In this plane, a region of parameter where the final fate converges to extremal BH is called the "extremal regime," and the other one in which it acts like trivial BHs with total evaporation called "decay regime."
The approximate transition line is given in Fig. \ref{fig_5}. Assuming $\alpha\gg 1$ gives the following analytical transition line,
\begin{equation}
\left. \frac{Q}{M}\right|_{\rm{Transition}} = \frac{8 \pi m M \alpha_{0}^{2}/\alpha}{1+ 8 \pi m M \alpha_{0}^{2}/\alpha^{2}},
\end{equation}
where for electron $\alpha_{0}=\frac{1}{\sqrt{4\pi \varepsilon_{0}G}}\frac{e}{m_{e}}=2\times10^{21}$.

Let us explore an illuminating property of the large $\alpha$ case. When $\alpha\gg1$, the geometry (\ref{eq6}) becomes flat at the extremal limit which is the final state of the evolution for any BH at extremal regime (Fig. \ref{fig_5}). Taking $r'=r-r_{-}$,
\begin{equation}
ds^{2}=dt^{2}-dr'^{2}-r'^{2}d\Omega ^{2}.\label{eq55}
\end{equation}
Therefore, it resembles an even more accurate description of an elementary particle \cite{Holzhey:1991bx}.

Taking the asymptotic behavior of the solutions to the wave equation for the particles (with angular parameter $n$) leaving the horizon in terms of the transition $T_{n}$ and reflection $R_{n}$ coefficients, and from  conservation of flux at the horizon and infinity we obtain
\begin{equation}
(\omega-\frac{qQ}{r_{+}})(1-|R_{n}(\omega)|^{2})=\omega |T_{n}(\omega)|^{2}.
\end{equation}
We see that for $qQ>0$, flux at infinity vanishes at a frequency due to the horizon electrical potential $\omega=q\Phi_{H}=\frac{qQ}{r_{+}}$. As stated previously these particles hardly escape the BH. Furthermore, low frequency modes encounter this hard electrical barrier at $r_{\Phi}=\frac{qQ}{\omega}$. We have observed this behavior in Fig. \ref{fig_1}. Owing to the large value of $\frac{e}{m}\simeq 2\times10^{21}$ for electrons, we can get incredibly large range for frequency and distance $r_{\Phi}$.
The greybody factors which are defined as the transition probability of waves passing through the BH potential for a given mode, are related to the reflection coefficient as $\gamma_{n}(\omega)=1-|R_{n}(\omega)|^{2}$.
We shall use this equation to show how dilaton BHs act like a firewall and produce echoes in the next section.


\section{Firewall behavior of dilaton black holes}
In the former section we have presented several unique features of dilaton BHs not seen in other types of BHs. Finally, in this section we end by bringing up several other nontrivial features  to establish the main scope of this paper.

We have the following differences between the traditional picture of a large mass BH and dilaton BHs:
\begin{enumerate}[label=(\alph*)]
    \item Curvature of spacetime near the event horizon is small for large mass BHs, while for dilaton BHs due to curvature singularity at $r_{-}$ \cite{Holzhey:1991bx,PhysRevD.43.3140} it is not always true.
    
    \item As seen in (\ref{eq2.14}) the area of the dilaton BH shrinks to zero size, and consequently its Bekenstein-Hawking entropies at the extremal limit, unlike Schwarzschild, RN, or Kerr BHs.
    
    \item Hawking temperature of the dilaton BH can blow up at the extremal limit for $\alpha>1$, unlike other types of BHs.
    
    \item We have a central curvature singularity at $r=0$ for the Schwartschild BH, while dilaton BHs possess two curvature singularities at $r=0$ and $r=r_{-}$.
    
    \item Unlike other types of BHs, this BH can evolve to extremal limit and become stable through the Hawking process as discussed in the former section and \cite{Abedi:2013xua}. This process can take place in finite time, not being in violation of the third law of BH thermodynamics (as the temperature blows up for $\alpha>1$ and is finite for $\alpha=1$ at this limit).
    
    \item As discussed in the former section and \cite{Abedi:2013xua}, for large $\alpha\gg1$ and the extremal limit, the geometry becomes flat. This appealing final state resembles even better the elementary particle description of these BHs.
\end{enumerate}

For both scalars and fermions we have seen the strong dependence on $\alpha$ with three distinct behaviors for $\alpha<1$, $\alpha=1$, and $\alpha>1$. Our range of interest is $\alpha\geq1$, where the peak of the potential barrier can get arbitrarily close to the horizon until it touches the event horizon at the extremal limit. These BHs mimic a firewall behavior. The peak of the potential barrier grows as $(r_{+}-r_{-})^{-2(\alpha^{2}-1)/(\alpha^{2}+1)}$ as one approaches the extremal limit. For the case of $\alpha=1$, the height of the potential barrier at the extremal limit remains finite, while for the class of BHs with $\alpha>1$ in this limit, it diverges on the event horizon. These BHs with $\alpha>1$ act like a perfect mirror at this limit. Interestingly the tortoise coordinate $r_{*}$ for the case of $\alpha>1$ at the extremal limit and at the event horizon is finite \cite{Abedi:2013xua}. One may wonder at having an exotic compact object with infinite compactness at this stage, where the barrier turns into a perfect one (infinite height). This unique feature occurs solely due to curvature singularity at $r_{-}$ which is inside the event horizon. This is an example of how a firewall inside the event horizon induces nontrivial properties to the outside, and how it creates an ECO.

Interestingly, a null ray that is trying to escape from the surface of this ECO does not need to be in a particular angle since the area of this ECO shrinks to zero at the extremal limit as seen in Eq. (\ref{eq2.14}) and \cite{Abedi:2013xua}. In another words there is no angular dependence for this metric (\ref{eq6}) at this limit. This is a main counterexample in this toy model to the argument given by Guo {\it et al}. 2017 \cite{Guo:2017jmi}. In another words, since the area of the dilaton BH shrinks to zero at the extremal limit, nothing can fall in. This example reveals how existing theories of quantum gravity may twist and turn from our trivial expectations. Since the Bekenstein-Hawking entropy of these BHs shrinks to zero at the extremal limit, one may wonder about presenting it as a featureless surface.

Here we provide two arguments pointing to the weakness of f nonobservable gravitational wave echoes to be as general brought by Guo {\it et al}. 2022 \cite{Guo:2022umn} via this toy model:

\begin{enumerate}[label=(\alph*)]
    \item Because of the overall mass and charge dependence of the peak of the barrier in (\ref{eq2.33}), the in-falling particles/fields hit the macroscopic reflective barrier before making a trapped surface. Therefore, the surface of ECO (barrier in the former section) remains  outside the trapped surface at least for the simple process of nearly spherical BH formations. Additionally, we always get an observable ringdown and quasi normal modes (QNM) spectrum from the barrier itself due to perturbations and partial reflection.
    \item Interestingly, although at the extremal limit and for $\alpha\geq1$, the peak of the barrier has touched the event horizon, it still has width (shown in Fig. \ref{fig_1}) which extends its reflective behavior far from the horizon.
\end{enumerate}

Let us bring up the same example raised by Guo {\it et al}. 2022 \cite{Guo:2022umn}: A BH of mass M that is created by N particles, each moving radially inward at the speed of light. No causal signal can travel from any of these particles to any other particle. Since the causality also holds in dilaton gravity, each particle must move exactly as it would if the other particles were not present. Considering a static hole, each particle definitely confronts the macroscopic near horizon geometrical barrier with $\alpha\geq1$ and runs into partial or total reflection (depending on the amount of charge they pose and value of the coupling constant $\alpha$) before reaching the radius $r=r_{+}$. For the case of total reflection no trapped surface and event horizon is formed. However, the particles shall follow the trajectory dictated by the geometry formed in dilaton gravity. In this case one may argue that this argument justifies the dynamical black holes described in \cite{Guo:2022umn}, given that this process of dilatonic black hole formation at the dynamical level may need numerical relativity computations. Consider an adiabatic approximation of spherically in-falling particles (wave packet). They possess charge $\delta Q$ and mass $\delta M$, so the exterior geometry of this wave packet changes (with $M\rightarrow M+\delta M$ and $Q\rightarrow Q+\delta Q$) with a trapped surface at $r_{+}+\delta r_{+}$ and maximum of the macroscopic effective potential barrier (\ref{eq2.33}) at $r_{\rm{max}}+\delta r_{\rm{max}}>r_{+}+\delta r_{+}$. Now we have two scenarios: One is for small $\delta Q$ and $\delta M$ where the trapped surface remains inside $r_{+}+\delta r_{+}<r_{\rm{max}}$. In this case the particles will reflect back at $r_{\rm{max}}$ before reaching the trapped surface at $r_{+}+\delta r_{+}$. Note that there is a similar argument in this scenario by Dailey et. al. 2023 \cite{Dailey:2023mvn} with numerical relativity justification for Schwarzschild black hole. For the other case (particles with higher  $\delta Q$ and $\delta M$) where the trapped surface moves to $r_{+}+\delta r_{+}>r_{\rm{max}}$, it becomes complicated to interpret the outcome without extensive numerical relativity justifications, and there might occur scenarios against our argument. However, in order to find a counterexample we can still bring up some cases that we could argue in support of this scenario. Consider the case of interest in this paper, the extremal or near extremal black holes with $\alpha>1$. In this example the black hole has zero or near zero area, with a curvature singularity at $r_{\rm{max}}=r_{+}$ or $r_{\rm{max}}\simeq r_{+}$ respectively. However, the wave packet (collapsing particles) is an extended object. The wavelength of the in-falling particles is still much larger than that of the black hole zero or near zero area even for smaller wavelengths. From a scattering point of view for the in-falling waves it might be hard to probe such small length scales. Let us assume $\delta r_{+}\sim \frac{1}{r_{+}}$ for this case with $r_{\rm{max}}=r_{+}$ or the trapped surface at $r_{+}+\delta r_{+}$ with $\delta r_{+}\ll r_{+}$. For these particles with potential barrier $V(r)$, using the barrier equation in \cite{Abedi:2013xua} for the waves with frequencies  $\omega\sim\frac{1}{r_{+}}$ the reflection occurs around $\omega^{2}\sim V(r_{\rm{{refl}}})$ at distance $r_{\rm{{refl}}}$ where we get $\frac{1}{r_{+}^{2}}\sim\frac{1}{r_{\rm{{refl}}}^{2}}(1-\frac{r_{+}}{{r_{\rm{{refl}}}}})^{\frac{2-2\alpha^{2}}{1+\alpha^{2}}}$ from the potential equation. For large $\alpha$ we find that $r_{\rm{{refl}}}=r_{+}+\mathcal{O}(r_{+})$, which is outside the trapped surface at $r_{+}+\delta r_{+}$.

In addition, as discussed in the former section dilaton BHs with $\alpha\gg1$ also contain the gently curved spacetime assumption by Guo {\it et al}. 2022 \cite{Guo:2022umn}.

\begin{figure}[t]
\centering
\includegraphics[width=0.48\textwidth]{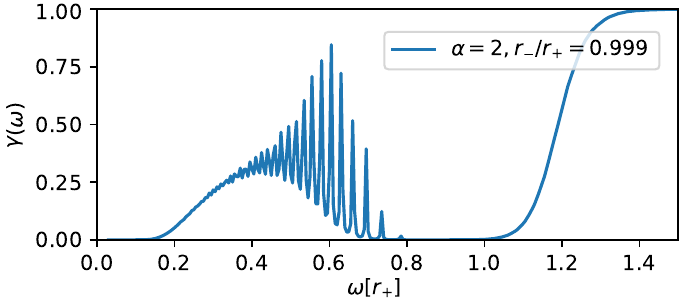}
\caption{\label{greybody}Greybody factors for $qQ>0$ at the near extremal limit (see the Appendix for details). As seen in this plot this BH replicates the ECO/firewall. These resonances at low frequency are due to repeating reflections or echoes from two barriers shown in Fig. \ref{fig_1}.}
\end{figure}

We also solve wave equations for fermions to find the greybody factors to bring an even more robust result showing how BHs in this gravity act like firewall producing echoes. As previously stated the peak of the potential barrier approaches the horizon and can get an indefinitely high value (growing as $(r_{+}-r_{-})^{-2(\alpha^{2}-1)/(\alpha^{2}+1)}$ at the extremal limit) depending on the value of $\alpha$ and the BH charge. The position of the peak for our interesting cases $\alpha\geq1$ at the extremal limit ($\epsilon\rightarrow0$) approaches the event horizon ($r_{\rm{max}}\rightarrow r_{+}$). For $\alpha>1$, the peak approaches as $r_{\rm{max}} \rightarrow \left(1+\frac{1}{2}\frac{\alpha^{2}+1}{\alpha^{2}-1}\epsilon\right)r_{+}$. This replicates an ECO/firewall behavior which is shown in Fig. \ref{fig_1}. Since the tortoise coordinate is finite for these types of BHs, the particles are reflected back in a finite time. For $\alpha=1$ this peak approaches as $r_{\rm{max}}\rightarrow\left(1+\sqrt{\frac{\epsilon}{2}}\right)r_{+}$ remaining finite when it touches the event horizon resembling a soft firewall. There is also another barrier discussed in the former section due to the electric potential of the BH. It reflects back low energy particles at distance $r_{\Phi}=\frac{qQ}{\omega}$ to the horizon, shown in Fig. \ref{fig_1}. So the existence of these two barriers makes a cavity producing multiple echoes as seen in Fig. \ref{greybody}, where the echoes in the time domain are presented as resonance harmonics in the frequency domain.


\section{Conclusion}
One may consider that the curvature singularity at $r=r_{-}$ breaks our semiclassical approximation near the event horizon when we are at the extremal limit. Accordingly, the quantum gravitational effects play a central role here and force the geometry to change significantly, not only for quantum gravitational effects, but also the vacuum expectation of the stress tensor $\langle T_{\mu\nu}\rangle$ for particles blow up as they grow with the curvature of spacetime as seen in \cite{Abedi:2015yga}. In this scenario we may expect that these effects due to vacuum fluctuations obstruct the singularity impeding the collapse to form a horizon \cite{Abedi:2015yga}. All these scenarios are leading to a firewall behavior. We may have a hot bath of particle creation acting again like a firewall. Since the Hawking temperature of these BHs with $\alpha>1$ blow up at the extremal limit, this is also consistent with our semiclassical picture as well. One may argue that these BHs radiate away their charge quickly through the Hawking radiation process, while it was shown by Abedi {\it et al}. 2013 \cite{Abedi:2013xua} that for some parameter space and initial state, the BH evolves toward the extremal limit (and the spacetime converges to a flat geometry for $\alpha\gg1$) when it reaches a stable extremal situation.

In this paper we brought BHs from the basic local action principle and show that we do not need to avoid the event horizon to get gravitational wave echoes. The singularity inside the event horizon changes the spacetime outside in such a way to the favor the firewall behavior and observable echoes. In particular, we  brought counterexamples from BHs in low energy limit of string theory $\alpha=1$, Kaluza-Klein theory $\alpha=\sqrt{3}$, and $\alpha\geq1$ in general, as our toy models answer the comments of Guo {\it et al}. 2017 and 2022 \cite{Guo:2017jmi,Guo:2022umn} about observable echoes without violating any basic assumptions.

\begin{acknowledgements}
We would like to thank Niayesh Afshordi for fruitful discussions and comments. J.A. was supported by the ROMFORSK Grant Project. No. 302640.
\end{acknowledgements}

\appendix*


\section{NATURAL UNITS AND NUMERICAL VALUES}\label{C}
The natural units and numerical values in plots are $G=\hbar=c=4\pi\varepsilon_{0}=1$, $r_{+}=100$, $q[r_{+}]=0.005$, $\omega[r_{+}]=0.6$, $qQ/\omega r_{+} =1.5$, $\frac{q}{m}=100$, $\kappa=1$, $\alpha_{0}=40$.

\bibliography{PRD}
\end{document}